# HIGHER SCHOOL OF ECONOMICS
### NATIONAL RESEARCH UNIVERSITY

*F. Aleskerov, V. Oleynik*

## MULTIDIMENSIONAL POLARIZATION INDEX AND ITS APPLICATION TO AN ANALYSIS OF THE RUSSIAN STATE DUMA (1994–2003)

Moscow
2016

Editors of the Series WP7
"Mathematical methods for decision making
in economics, business and politics"
*Aleskerov Fuad, Mirkin Boris, Podinovskiy Vladislav*

    The multidimensional extension of the Aleskerov-Golubenko polarization index is developed. Several versions of the polarization index are proposed based on different distance functions. Basic properties of the index are examined. The behavior of polarization index is studied numerically in the case of the "uniform" distribution of groups in the unit square and unit three-dimensional cube.
    Polarization in the Russian State Duma (1994–2003) is studied using the index elaborated. The two-dimensional model of the Russian State Duma previously developed on the basis of publically available roll-call data is used in the analysis. The results of application of the multidimensional index are consistent with occurring political events. It is shown that polarization in the State Duma was mostly associated with the degree of tension in its relations with the executive authorities. Particularly, the more severe was the confrontation between the legislative and the executive branches of power, the less polarized was the State Duma, and vice versa

# Table of Contents

# 1. Introduction

Social polarization has recently become at the focus of interest of researchers from a broad range of fields. As any social concept, social polarization may be defined in a number of ways but the definition given by Esteban and Ray [1994] is now treated as classical and will be used as basic in the present work:

*"Suppose that a population of individuals may be grouped according to some vector of characteristics into clusters, such that each cluster is very similar in terms of the attributes of its members, but different clusters have members with very dissimilar attributes. In that case, we say that the society is polarized".*

Social polarization defined as above, is closely related to an antagonism in society and its radical manifestations. There are many studies emphasizing the inextricable link between polarization and conflicts. In his seminal work Horowitz [1985] states that in societies where a large ethnic majority is present, when a large enough ethnic minority faces the former, the most sever conflicts arise. Numerous empirical studies confirm the connection between polarization and conflicts. Particularly, it is shown that polarization index based on income is a significant predictor of social conflicts (see [Esteban et al., 2007]), ethnic polarization index is a predictor of civil wars (see [Montalvo, Reynal-Querol, 2005]). Moreover, not only empirical studies but also game-theoretic ones provide convincing evidence of link between polarization and conflict. Specifically, Esteban and Ray [2008] found connection between polarization index and equilibrium in a behavioral model of conflict; Montalvo and Reynal-Querol [2005] revealed connection between their polarization index and rent-seeking game.

Therefore, the motivation to conceptualize, measure and analyze polarization in society can hardly be overestimated.

### Acknowledgments

The authors are grateful to Vyacheslav Yakuba for the courtesy of providing data on the two-dimensional model of the Russian State Duma (1994–2003). The article was prepared within the framework of the Basic Research Program at the National Research University Higher School of Economics (HSE) and supported within the framework of a subsidy by the Russian Academic Excellence Project "5–100".

## 1.1. A Review of the Literature

During the last two decades a solid corpus of literature emerged dealt with the analysis of polarization in societies and, especially, to the approaches of social polarization measurement. The pioneering works by Esteban and Ray [1991, 1994] and Wolfson [1994], in fact, outlined two main directions in the measurement of social polarization.

According to the first framework suggested by Esteban and Ray [1991, 1994], polarization is considered in terms of intra-group identification and inter-group alienation. Esteban and Ray, themselves, call their approach *identification-alienation* framework. Within this framework it is assumed that an individual in society identifies herself with a particular group and feels alienated from other groups. The more people feel strongly connected with *their* group and distant from *other* groups, the more society is polarized. In addition, the number and size of groups also matters. Thus, within the identification-alienation framework polarization rises if the groups become more homogeneous internally, more separated externally, and more equal in size. Moreover, if the number of groups is more than one, then the larger the number of groups, the lower the polarization.

Esteban and Ray take special care to the distinction between polarization and inequality. One of the reasons why they emphasize this issue is the enhanced attention that has been placed in the literature to inequality[1] against the backdrop of much less attention paid to polarization which is, from the authors' point of view, much more responsible for social tensions, conflicts and radical forms of their manifestation.

In this respect, let us present one striking example which illustrates how inequality differs from polarization. Consider a society with two distinct groups of people characterized by their income; namely, the low- and high-income population. Suppose, this year the variation of income of both low-and high-income people dropped comparing to the previous year, so that income distributions of these groups became denser. Evidentially, comparing to the previous year, inequality in the society decreased; however, polarization in this society raised.

---

[1] Even a separate field called *economics of inequality* was developed (see, e.g. [Foster, Wolfson, 1994; Esteban et al., 2007; Atkinson, 2003]).

By imposing a set of reasonable axioms, Esteban and Ray [1991, 1994] narrowed down the class of possible polarization measures to the following form:

$$ER = k \sum_{i=1}^{n} \sum_{j=1}^{n} \pi_i^{1+\alpha} \pi_j |y_i - y_j|$$

for some positive constants $k$ and $\alpha \in (0; \alpha^*]$, usually it is accepted $\alpha^* \approx 1.6$. Here, $y_i$ corresponds to income level of the group $i$ and $\pi_i$ is its share with respect to the whole society. The parameter $\alpha$ can be interpreted as the degree of "polarization sensitivity".

The strong connection between Esteban-Ray and Gini index may be revealed. Specifically, when $\alpha = 0$ and $k = 1$, $ER$ index is precisely the Gini coefficient.

It should be noted that $ER$ index with a discrete metrics $\delta(y_i, y_j) = |y_i - y_j|$, $\delta(y_i, y_j) = \begin{cases} 0 \; if \; i = j \\ 1 \; otherwise \end{cases}$ is known as Reynal-Querol index which was presented in Reynal-Querol [2002] and actively used in ethnic polarization studies.

Several other alternatives and modifications of the ER index were proposed. Particularly, identification-alienation approach was actively developed, among others, by Gradin [2000], Zhang and Kanbur [2001], Duclos et al. [2004], Esteban et al. [2007].

Another governing approach to the measurement of polarization in societies was mapped out by Wolfson [1994]. In this framework polarization is concerned with dispersion of the income distribution from the median (or alternatively defined center of the distribution) towards the extreme points. The decline of the middle class is described, measuring how the center of the distribution of the specific characteristics (income, education, etc.) is emptied. This approach is often called as "bi-polarization".

The Wolfson index of polarization can be defined as a function of the inequality between groups and the inequality within groups. Generalizing this approach, one can obtain a class of polarization measures defined as

$$P(\vec{x}, k) = f\big(I^W(\vec{x}), I^B(\vec{x}), S(k)\big),$$

where $f$ is a monotonic function in all arguments, $\vec{x} = (x_1, \dots, x_n)$ is a vector of a particular feature (income, political ideology, etc.) describing given society of size $n$ separated into $k$ groups, $I^W(\vec{x})$ and $I^B(\vec{x})$ are some measures of inequality within and between groups, respectively, and $S(k)$ is a measure of the $k$ group's size.

Among followers of the "bi-polarization" approach Wolfson [1997], Wang and Tsui [2000], Chakravarty and Majumder [2001], Rodriguez and Salas [2003], Chakravarty et al. [2007], Chakravarty and Ambrosio [2010], Gigilarano et al. [2011] can be mentioned.

The majority of the studies devoted to the measurement of polarization in societies considered only the case of the one-dimensional measures. Particularly, it was assumed that population in society is split into groups according to a single characteristic; in most cases, income level. However, societies are much more complex in their structure, and disagreement often arises over multiple issues. Thus, it is necessary to possess the techniques which allow to measure polarization in societies in which groups are formed according to multiple characteristics.

Multidimensional polarization measures have been of growing concern within last years. Notwithstanding moderate success achieved in this field, some studies of multidimensional polarization can be found in the literature.

Gigliarano and Mosler [2009] argue that splitting of the population can be based on multiple attributes such as education, wealth or health. They constructed a class of multidimensional polarization measures by decomposing different inequality measures with measuring the relative group size. According to them polarization consists of inequality within groups, and inequality between the groups given a sufficient group size. In essence, this measure is a multidimensional extension of the group approach of Esteban and Ray [1991, 1994].

Sheicher [2010] extends his own unidimensional polarization measure to the multidimensional index which is a combination of poverty and affluence measures. It is based on the distance of the income of a middle class individual to specific middle class thresholds.

Technically, multidimensionality in polarization measures respect multiple attributes via an attribute matrix in a grouping approach or have respecting multiple attributes via distances to the pole thresholds. Further developments are discussed in poverty analyses [Bossert, Chakravarty, D'Ambrosio, 2013; Nolan, Whelan, 2007; Atkinson, 2003].

In Poole and Rosenthal [1984] an approach is proposed to estimate polarization in the United States Congress. This approach is based on the use of well-known NOMINATE scores originally elaborated by Poole and Rosenthal [1983] which represent two-dimensional coordinates of the Congressmen in the latent political space. The first dimension is interpreted as "liberal – conservative" dimension, while the second dimension picks up regional dissimilarities, attitudes on cross-cutting, salient issues of the day (e.g., slavery, civil rights, lifestyle issues, etc.) (see [Poole, Rosenthal, 1997]).

Poole and Rosenthal proposed to measure polarization as distance between the Republicans and the Democrats means of the first dimension coordinate. In McCarty et al. [2006] the same approach was applied for newly developed joint scale DW-NOMINATE scores. The most recent results of estimation of polarization in the U.S. Congress are publically available on *Voteview* website.

The major shortcoming of the approach suggested by K. Poole and H. Rosenthal is the fact that by averaging the coordinates of the Congressmen we are losing the information about how scattered or dense are the clusters representing the Legislators, which, evidentially, has an impact on polarization in the Congress. On the other hand, the present multidimensional index based on the central moment of forces lacks such a flaw.

## 2. The Model

The following model is an extension of the one-dimensional model presented in Aleskerov and Golubenko [2003] to the multidimensional case.

To the best of our knowledge, Aleskerov-Golubenko polarization index (AG-index) is the first attempt to adopt physical concepts to the measurement of polarization in society. It was inspired by the notion of central moment of a system of forces coming from physics. The framework suggested by Alesrekov and Golubenko is new and alternative to two governing trends proposed so far; nevertheless, the measures developed within this framework satisfy reasonable requirements originally specified by Esteban and Ray [1994].

Suppose, we are given a society split into *n* groups according to particular criteria. Following the definition of social polarization by Esteban and Ray [1994] cited above, assume there exist *attributes*, or *features* of society that create similarities and differences between individuals, and, consequently, groups of individuals. Since we are considering multidimensional case, a vector of characteristics according to which individuals are grouped into clusters is multidimensional.

Each group is described by a number $v_i, i = \overline{1,n}$, and a vector $\overrightarrow{p_i} = (p_{i1}, \ldots, p_{im}), i = \overline{1,n}$, in a multidimensional space, $\mathbb{R}^m$, where $m$ is a number of attributes describing the society under consideration. A number $v_i, i = \overline{1,n}$, corresponds to the share of group's members in the whole society, $\sum_{i=1}^{n} v_i = 1$. For the sake of simplicity, thereafter we assume that each of $m$ attributes can be presented as a value of some scalar variable taking value from the interval $[0;1]$. For instance, consider ideological "leftists – rightists" scale where 0 corresponds to extreme leftists and 1 – to extreme rightists, respectively. Thus, $\overrightarrow{p_i}, i = \overline{1,n}$, is a point in the multidimensional be $[0;1]^m$ representing positions of the group in respective dimensions.

Hence, each group may be seen as a *weighted point* in the multidimensional unit cube; together all the groups form a system of weighted points.

Let us now define a *center of mass* $\vec{c} = (c_1, c_2, \ldots, c_m)$ of the system of points $\overrightarrow{p_i}, \ i = \overline{1,n}$, in which weights $v_i, i = \overline{1,n}$, are concentrated respectively, as

$$\vec{c} = \frac{\sum_{i=1}^{n} v_i \overrightarrow{p_i}}{\sum_{i=1}^{n} v_i} = \sum_{i=1}^{n} v_i \overrightarrow{p_i}$$

Then, the multidimensional polarization index of a society under consideration may be written in the following way

$$P = k \sum_{i=1}^{n} v_i \cdot d(\overrightarrow{p_i}, \vec{c}),$$

where $d \colon \mathbb{R}^m \times \mathbb{R}^m \to \mathbb{R}$ is some distance function and $k$ is a normalizing coefficient. For definiteness let us present three versions of polarization index with respect to different forms of metric $d$.

If $d$ is defined as Euclidean metric, then polarization index takes a form

$$P_{Euc} = \frac{2}{\sqrt{m}} \sum_{i=1}^{n} v_i \cdot \sqrt{\sum_{j=1}^{m} \left(p_{ij} - c_j\right)^2}.$$

If $d$ is defined as Manhattan distance, then polarization index takes a form

$$P_{Man} = \frac{2}{m} \sum_{i=1}^{n} v_i \cdot \sum_{j=1}^{m} |p_{ij} - c_j|.$$

If $d$ is defined as Chebychev distance, then polarization index takes a form

$$P_{Cheb} = 2 \sum_{i=1}^{n} v_i \cdot \max_{j=\overline{1,m}} |p_{ij} - c_j|.$$

The normalizing coefficient $k$ is selected in such a way that the maximal value of $P$ index was equal to 1.

In Lipacheva [2015], *inter alia*, the properties of unidimensional Aleskerov – Golubenko polarization index have been studied. In particular, asymptotical properties of the index have been examined. It has been shown that if $n \to \infty$ $P \to \frac{1}{2}$. However, it was proposed that polarization index should shrink to zero as the number of groups in society tends to infinity. Thus, a slight modification of this index was presented. In the multidimen-sional case the modified version of index can be defined in the following way

$$P' = \frac{k'}{n}\sum\nolimits_{i=1}^{n} v_i \cdot d(\overrightarrow{p_i}, \vec{c}).$$

Different versions of $P$ can be suggested according to different types of metrics:

$$P'_{Euc} = \frac{4}{n\sqrt{m}}\sum\nolimits_{i=1}^{n} v_i \cdot \sqrt{\sum\nolimits_{j=1}^{m}(p_{ij} - c_j)^2},$$

$$P'_{Man} = \frac{4}{n \cdot m}\sum\nolimits_{i=1}^{n} v_i \cdot \sum\nolimits_{j=1}^{m}|p_{ij} - c_j|,$$

$$P'_{Cheb} = \frac{4}{n}\sum\nolimits_{i=1}^{n} v_i \cdot \max_{j=\overline{1,m}}|p_{ij} - c_j|.$$

In this form, $P'$ index turns to zero in case of a single group and shrinks to very low values if the number of groups in society is very high.

Nevertheless, it seems to us that even if the number of groups is large, there still might be some tension within society, and polarization does not shrink to zero. In addition, larger values of polarization index simplify the comparison of results and overall analysis. Therefore, in empirical part of our research we decided to use the original form of the $P$ index.

## 2.1. Basic Properties of the P Index

The maximum of $P$ index equals 1 (due to normalizing coefficient) and is attained in the society divided into two equally-sized groups which are positioned in two extreme points of the main diagonal of a unit cube. For instance, in two-dimensional case there are four possible configurations when maximal value. Namely, when one of two equal groups is placed at $(0, 0)$ and the other at $(1, 1)$, or when one group's coordinates are $(0, 1)$, while the other's equal $(1, 0)$. However, this maximum is not unique. In what follows, other configuration delivering maximum will be examined.

$P$ index takes only non-negative values since values of distance function $d(\cdot)$ and shares $v_i, i = \overline{1, n}$, are non-negative.

The minimum of $P$ index is equal to zero and is attained when all groups in the society are positioned at the same point, or when the unique group represents the whole society. The center of mass of the system of points coincides with the position of all groups in the first case and the position of the single group in the second case; thus, whatever form of distance function has been chosen, the polarization index takes zero value.

Now, let us consider a special case of "equally" ("uniformly") distributed groups.

Suppose, a society is divided into $n$ equal groups which are placed in a unite cube in such a way that any pair of neighboring groups are equally distanced. In other words, these groups are placed at uniform grid points, so that each of their coordinates takes values $\frac{i-1}{n-1}, i = \overline{1, n}$. Evidentially, in such a case number of groups is equal $l^m$, where $m$ is the space dimensionality and $l$ is some positive integer.

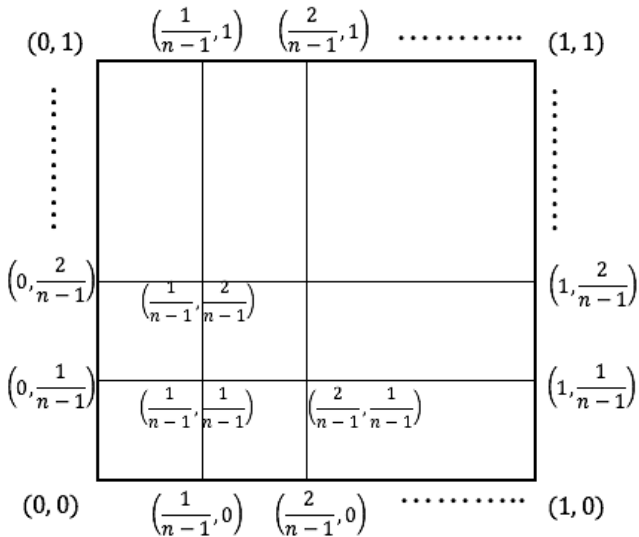

**Fig. 1.** "Equal" distribution of n groups in two-dimensional case

Even though such a distribution is fairly artificial and not likely to be observed in reality, it is still interesting to be examined.

Since analytical study of this special distribution even for two-dimensional case is somewhat inelegant and is associated with cumbersome calculations, it was analyzed numerically.

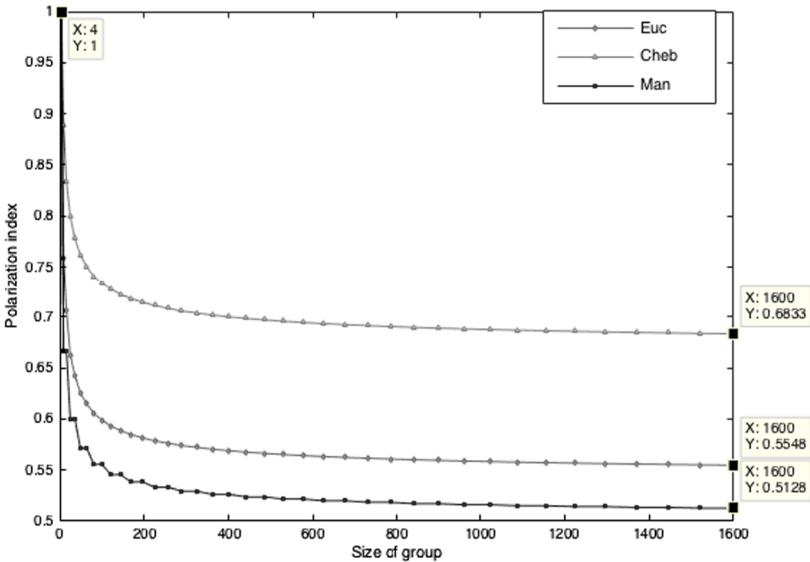

**Fig. 2.** Behavior of polarization indices in the case
of "equal" distribution of groups in two-dimensional space

As Fig. 2 shows, in the case of groups equally distributed in two-dimensional case, polarization index based on Chebychev distance is always not less than both indices based on Euclidean and Manhattan distances. At the same time, polarization index based on Manhattan distance was always not larger than two other indices. All these indices attain maximal value equal 1 when the number of groups is four (therefore, as it was mentioned above, maximum of the polarization index is not unique). Each version of polarization index converges and monotonically decreases while the number of groups grows.

Analogously, Fig. 3 demonstrates a behavior of polarization indices in the case of "equal" distribution in three-dimensional feature space. Again, polarization indices monotonically decrease when the number of group increases, and all of them converge to respective values.

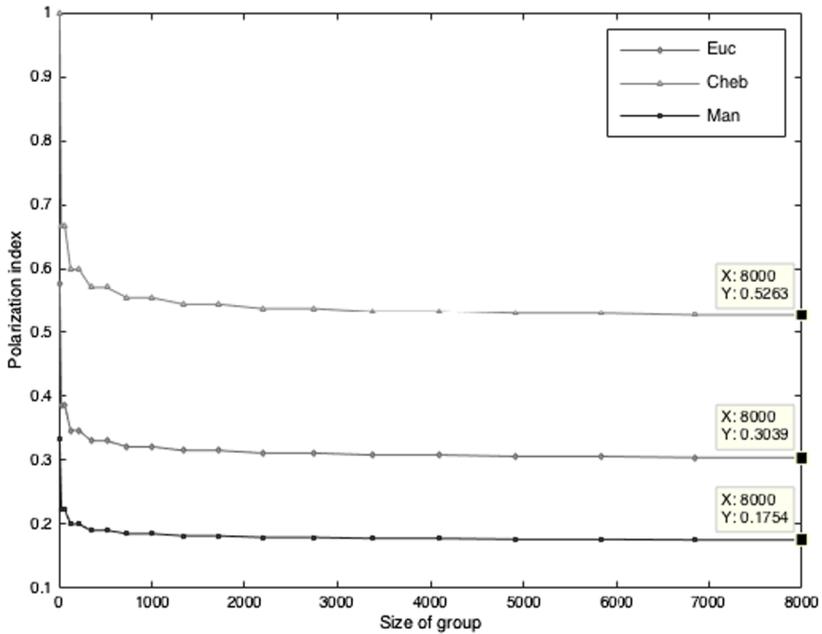

**Fig. 3.** Behavior of polarization indices in the case
of "equal" distribution of groups in three-dimensional space

## 3. Empirical Applications

### 3.1. Case of the State Duma of the Russian Federation (1994–2003)

Let us now apply the methodology elaborated to an analysis of the Russian State Duma (1994–2003).

The State Duma is a lower house of the bicameral parliament, the Federal Assembly, which is the national legislature body of the Russian Federation. The upper house is called the Federation Council.

The State Duma had a thorny destiny and tangled history. The first elected parliament was introduced in 1906 during the reign of Tsar Nicholas II.

However, the first two Dumas were extremely ineffective and were quickly dissolved. Subsequent convocations of the State Duma were more viable than the first attempts; however, Russian parliament was one of the factors that contributed to the eradication of autocracy in the Russian Empire. Finally, in 1917, after the February Revolution the State Duma was abolished. Later on, from 1938 until 1993 the Supreme Soviet of the Russian SFSR was the supreme government body performing legislative functions. This parliament was forcibly dissolved by President Yeltsin. According to the new Constitution introduced by President Yeltsin after constitutional crisis of 1993, the State Duma became the new generally elected legislative institution of the Russian Federation. According to the new Constitution, the State Duma of the Russian Federation is elected for a term of four years, and the procedure of the elections of deputies is determined by federal laws. The Russian State Duma was reelected 6 times: in 1993, 1995, 1999, 2003, 2007, and 2011.

The State Duma consists of 450 deputies. Until the fall 2005, elections were held under a mixed system. 225 of deputies were elected in nationwide electoral districts proportionally to the number of votes cast for the federal lists of candidates nominated by electoral parties, the remaining 225 deputies were elected in single-mandate electoral districts. Since July, 2005 all of 450 deputies are elected under a proportional system. In 1993–2003 electoral threshold was set at 5% level, but since the parliamentary elections of 2007 this threshold is equal to 7%.

The deputies have a constitutional right to form so called deputy unions: factions and deputy groups. Each deputy can be a member only of a single deputy union. Factions are organized according to the party membership (only parties elected in nationwide federal district have a right to from factions); deputies not being members of factions can form deputy groups according to their preferences. The minimal number required to from a deputy group was equal to 35 deputies until 2004; afterwards this number accounted for 55 deputies.

In the State Duma federal laws are adopted by simple majority rule (i.e., 226 votes are needed to pass a federal law) while constitutional laws are adopted by qualified majority rule (namely, 2/3 out of 450 votes, i.e., 300 votes are needed to adopt such laws).

In this paper we consider three first convocations of the State Duma which took place during 1994–2003. The publically available roll-call votes of deputies provided by INDEM-Statistics enabled Aleskerov et al. [2007] to apply an algorithm originally suggested in Poole and Rosenthal [1983] and thoroughly described in Satarov [1993] in order to build spatial models, or so called *political maps* reflecting political positions of deputies of the Russian State Duma. On the whole, the problem of political maps building boils down to three main sub-problems: finding principal components (latent factors of political disengagement), estimation of so called factor loadings and calculation of corresponding values of factor variables for deputies-objects. A procedure suggested in Satarov [1993] and described in detail in Blagoveshchensky [2004] consistently solves this problem. Firstly, latent factors of political disengagement are determined by means of multidimensional non-metric scaling; points obtained are centered and oriented along principal components. Then, for every vote the parameters are calculated determining to which extent the split in positions during the voting is caused by the contribution of each factor. Finally, "benchmark" votes are defined according to the factors determined, and taking into account these parameters coordinates of all deputies are calculated – kind of ratings of compliance with "benchmark" votes on each of the factors of political disengagement. Thus, this three-step procedure enables to build a political map of the parliament in which every deputy is represented as a single point in a multidimensional space.[2]

It turned out that positions of deputies in the Russian parliament in 1994–2003 could be presented in a two-dimensional space. According to the interpretation given in Aleskerov et al. [2007], the informative meaning of the latent factors determined be means of the aforementioned algorithm are "Loyalty – Opposition" to the current executive authorities, on the one hand, and "Ideology – Pragmatism" (during the 1st convocation) and "Liberals – Statesmen" (during the 2nd and the 3rd convocations), on the other hand.

Originally, in our model we consider positions of groups rather than individuals, so we need to specify the coordinates of groups of deputies.

---

[2] Usually, linear transformations are applied to the points representing deputies, so that their coordinates lie in the multidimensional unit cube.

We will assume that coordinates of a particular group are defined by the average coordinates of all the deputies belonging to this group as it was done in Aleskerov et al. [2007].

*The State Duma of the 1$^{st}$ convocation (1994–1995)*

Elections of the deputies of the State Duma of the 1$^{st}$ convocation were held under a mixed electoral system. 225 deputies were elected in nationwide federal district and 219 deputies – in single mandate districts. Elections were not fulfilled in 5 electoral districts and were not held in Chechen Republic at all. 4 parties managed to cross the electoral threshold of 5%: "Liberal Democratic Party of Russia" (an unexpected favorite in the federal district), "Democratic Party of Russia", "Women of Russia", and "Party of Russian Unity and Accord". "Russia's Choice" won fewer votes in the federal district than expected and managed to form the largest deputy union in the State Duma of the 1$^{st}$ convocation only due to the deputies elected in single-mandate districts. Deputy unions formed immediately after elections have been changing during the term of the 1$^{st}$ convocation. We will consider the following 11 deputy unions present in the State Duma of the 1$^{st}$ convocation.

– "Agrarian Party of Russia" (APR)
– 'Russia's Choice" (RC)
– "Democratic Party of Russia" (DPR)
– "Women of Russia" (WR)
– "Communist Party of the Russian Federation" (CPRF)
– "Liberal Democratic Party of Russia" (LDPR)
– "New Regional Politics" (Duma-96)
– "Party of Russian Unity and Accord" (PRES)
– "Liberal Democratic Union of 12 December" (LDU) --
– "Russia" (RUS)
– "Stability" (STAB)
– "Yabloko" (YAB)

There were deputies who were not affiliated to any deputy union. For these independent deputies, in Aleskerov et al. (2007) additional classification was given.

– Independent LDU – the deputies who were members of the deputy group LDU

- Independent "Democracy" – the deputies who were not affiliated to deputy unions and who later on belonged to the deputy group "Democracy" close to CPRF
- Independent "Outcasts" – well-known deputies who had specific reputation
- Independent – all other deputies who were not affiliated to any formal or informal groups of deputies

In Aleskerov et al. [2007] spatial model of the State Duma of the 1[st] convocation was built through exploitation of roll-call votes data and three-step algorithm described above. Fig. 4 shows political map of the State Duma in 1994.

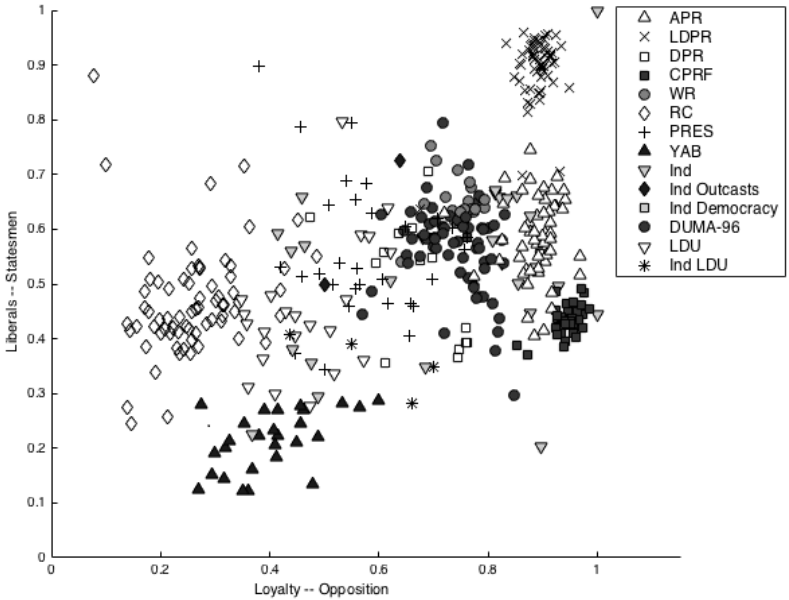

**Fig. 4.** Spatial model of the State Duma in 1994

Since the number of independent deputies not affiliated to any formal or informal group was quite large, we suggest an additional technique which enables to attach these deputies to a particular deputy union according to *"nearest neighborhood"* rule. Specifically, for each independent deputy we consider some vicinity (of a particular radius) and three nearest neighbors in this vicinity.

We attach an independent deputy to the deputy union to which two out of three nearest neighbors belong. If there is no such a union, the deputy is treated to be independent. Such an approach is based on the idea that the position of the independent deputy is represented by his coordinates in the spatial model of the parliament. Thus, independent deputies who are close to the position of their colleagues belonging to a particular faction or deputy group, reveal voting patterns similar to this union; therefore, they most probably hold the opinions similar to the opinions of this deputy union.

The independent deputies who were not attached to any formal or informal group are treated as a separate united cluster (Fig. 5). Fig. 6 – 7 represent the positions of the deputy unions and groups of independent deputies in the State Duma in 1994.[3] The size of the marker corresponds to the deputy union's weight (the share of the seats the deputy union possesses in the parliament).

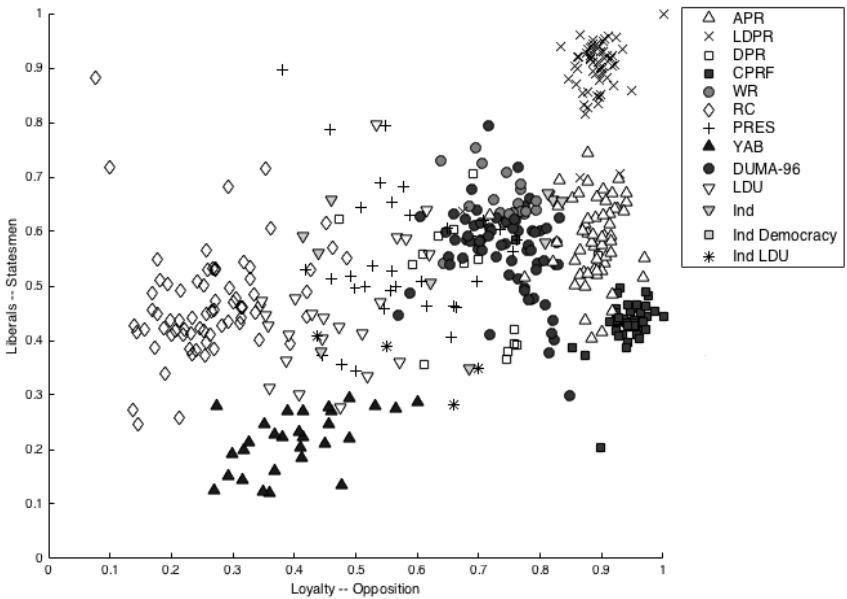

**Fig. 5.** Spatial model of the State Duma in 1994. Independent deputies are attached to the deputy union according to the "nearest neighborhood" rule

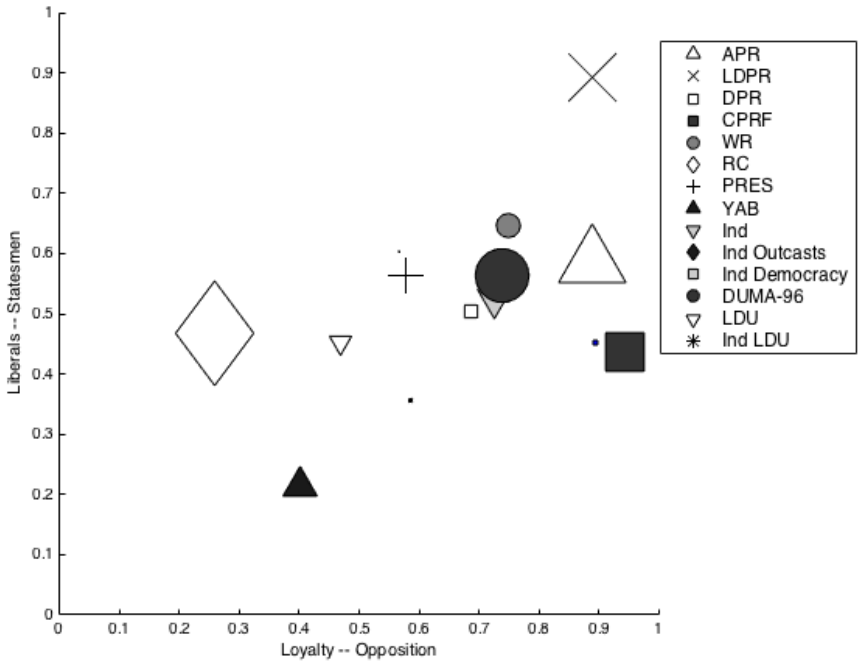

**Fig. 6.** Positions of the deputy unions in the State Duma in 1994

One of the specific features of the State Duma in 1994 was lack of rigid discipline within deputy factions and groups. Fig. 4 clearly demonstrates this specific feature: clusters representing deputy unions are on average rather sparse than dense. However, particular deputy unions can be characterized by stronger discipline than others, and their clusters are denser on the political map. Specifically, all deputy factions, namely, CPRF, LDPR, WR, RC reveal quite high solidarity in their members' positions. At the same time, deputy groups are represented by much more disperse clusters. Only deputies belonging to the union "Yabloko" have quite common positions on the political map.

Cross-factional migrations of deputies were not uncommon in the State Duma in 1994. Uncertainty of the positions of deputy unions left to deputies the room for maneuvers. In general, configuration of deputy unions in the State Duma reflects the current moods at that time: pluralism appeared in the Russian politics. As Fig. 6–7 demonstrate, there were several quite numerous deputy unions having different positions (deputy unions are scattered over almost the whole unit square), and none of these unions had overwhelming advantages.

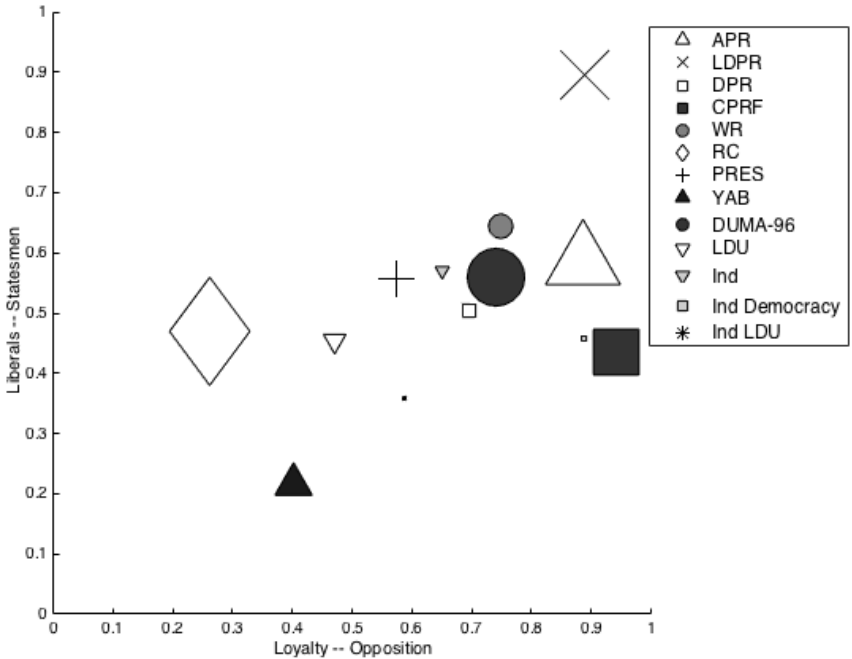

**Fig. 7.** Positions of the deputy unions in the State Duma in 1994. Independent deputies are attached to the deputy union according to the "nearest neighborhood" rule

Now, when both shares and coordinates of the deputies' unions are determined, we are able to calculate $P$ indices.

The results obtained are presented below.

$Center\ of\ mass = (0.6746, 0.5523);$
$P_{Euc} = 0.3479, P_{Man} = 0.3136, P_{Cheb} = 0.4487.$

Now, let us calculate $P$ index in case when independent deputies are attached to the deputy unions according to the "nearest neighborhood" rule.

$Center\ of\ mass = (0.6746, 0.5523);$
$P_{Euc} = 0.3571, P_{Man} = 0.3212, P_{Cheb} = 0.4607.$

As we see, the center of mass, evidently, did not change. At the same time, the value of $P$ index (under different forms of metrics) almost did not change.

The coordinates of the center of mass show that on average, deputy unions had quite moderate positions along "Ideology – Pragmatism" dimension. However, it can be concluded that the State Duma on the whole was more opposed than loyal to the current executive authorities. Several events which have taken place at that time support this conclusion. Firstly, I. Rybkin, the member of the opposition "Agrarian Party of Russia", was elected as a Chairman of the State Duma. Furthermore, the State Duma voted for the Amnesty for participants of the events of 1991 and 1993; subsequently, opponents of President Yeltsin were released.

Values of polarization indices are moderate. The positions of deputy unions were not close enough which made the parliament quite polarized; nevertheless, were they sparse enough to make the parliament quite balanced.

Now, let us present the analogous findings for the case of the State Duma in 1995.

The values of polarization index:

$Center\ of\ mass = (0.7668, 0.7231);$
$P_{Euc} = 0.2190, P_{Man} = 0.1876, P_{Cheb} = 0.2919$

Application of the "nearest neighborhood" approach to treatment of independent deputies almost did not change the view of political map. Thus, we do not provide the figures of political map and positions of the factions. However, we do provide the values of $P$ index which slightly increased:
$P_{Euc} = 0.2227, P_{Man} = 0.1909, P_{Cheb} = 0.22968.$

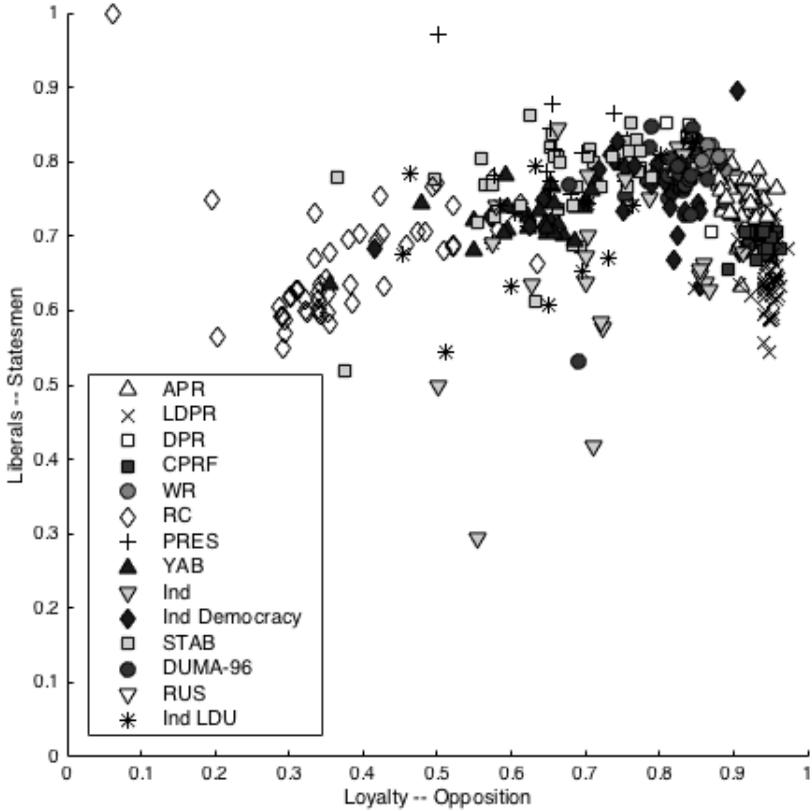

**Fig. 8.** Spatial model of the State Duma in 1995

Both points configuration on the political map and values of polarization indices clearly show that deputies were much more unanimous in their political positions in 1995 than a year before. Fig. 8–9 clearly indicate the drift of all deputy unions to the upper right area of the political map – actually, almost all points are concentrated in this quadrant of the unit square. Consequently, the center of mass has significantly shifted upwards and to the right comparing to the previous year (compare $(0.6746, 0.5523)$ in 1994 and $(0.7668, 0.7231)$ in 1995). Deputies became more opposed to the current executive authorities (first of all, represented by the President) and revealed more pragmatism while voting. At the same time, relatively high unanimity in deputies' positions on the political map is explicitly reflected in dropped values of polarization indices: they decreased about 1.5 times (within both groups-points and deputies-points approaches and for different distance functions) comparing to the case of 1994.

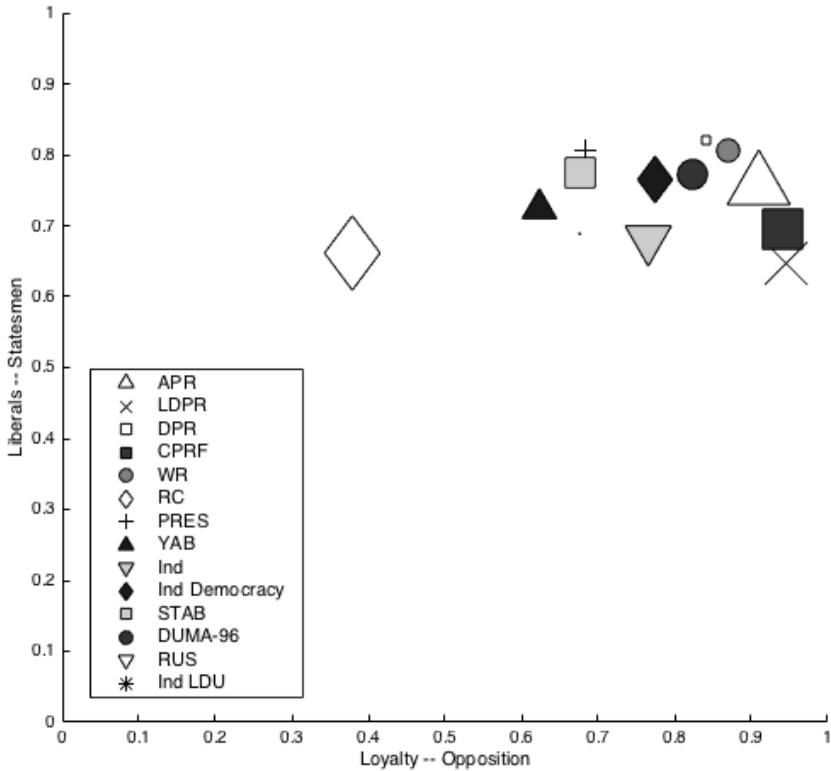

**Fig. 9.** Positions of the deputy unions in the State Duma in 1995

The results obtained exhibit a good agreement with political realities and, particularly, moods in the State Duma in 1995. One of the crucial events which gave strong impetus to changes of the political positions of deputies and contributed to the increased opposition of the State Duma on the whole was the launch of military operations in Chechnya. In addition, deputies exhibited quite cohesive position (again, opposed to the current authorities) while voting on the issue of confidence in the Government. In June, 1995, in connection with hospital hostage crisis in Budyonnovsk, the State Duma expressed non-confidence in the Government. A vote was blocked in the second ballot only after additional exhausting consultations and resignation of some power agents (so called "siloviki").

Thus, the State Duma in 1995 turned out to be much less polarized than in 1995; however, this lower polarization was not the result of positively oriented unanimity among deputies, but instead the byproduct of their confrontation with common enemy through the executive authorities.

## *The State Duma of the $2^{nd}$ convocation (1996–1999)*

According to the results of national elections held in December 1995, four parties cleared electoral threshold: CPRF, LDPR, "Our Home – Russia" and "Yabloko". Communists and their allies controlled 220 mandates in the parliament; hence, the $2^{nd}$ convocation of the State Duma turned out to be strongly influenced by the leftists opposed to the President and the Government. Besides four factions formed by parties which passed to the State Duma, three deputy groups were created: "Agrarian Group", "People's Power" and "Regions of Russia".

We will consider 7 following deputy unions present in the State Duma of the $2^{nd}$ convocation.

- "Agrarian Group" (AG)
- "Communist Party of the Russian Federation" (CPRF)
- "Liberal Democratic Party of Russia" (LDPR)
- "People's Power" (PP)
- "Our Home – Russia" (OHR)
- "Regions of Russia" (RR)
- "Yabloko" (YAB)

Analogously to the case of the $1^{st}$ convocation, the attempt was made to classify the independent deputies.

- Independent (SPS)[4] – the independent deputies characterized by democratic orientation (the majority of these deputies were elected to the State Duma of the $3^{rd}$ convocation and formed the faction SPS),

---

[4] SPS – from Russian «СПС», Союз Правых Сил, the Union of Right Forces.

– Independent (KREM) – the independent deputies who were ready to cooperate "constructively" with Kremlin but distance themselves from it.

As in the State Duma of the 1st convocation, two dimensions were found to be enough to describe its political space (see [Aleskerov et al., 2007]). The latent factor corresponding to the variation of the X-coordinate of deputies' positions is, again, called "Loyalty – Opposition" towards authorities. At the same time, the Y-coordinate latent factor can be called "Liberals – Statesmen" (in the case of foreign affairs issues, "Westerners – Patriots"). In principle, latent factor suggested for the case of the 1st convocation ("Ideology – Pragmatism") may be used as well; nevertheless, some biases may arise because of the large number of the communist deputies.

Fig. 10 represents the political map of the State Duma in 1996. We did not implement the "nearest neighborhood" technique for this case since the number of independent deputies who did not belong to any formal of informal group was relatively low.

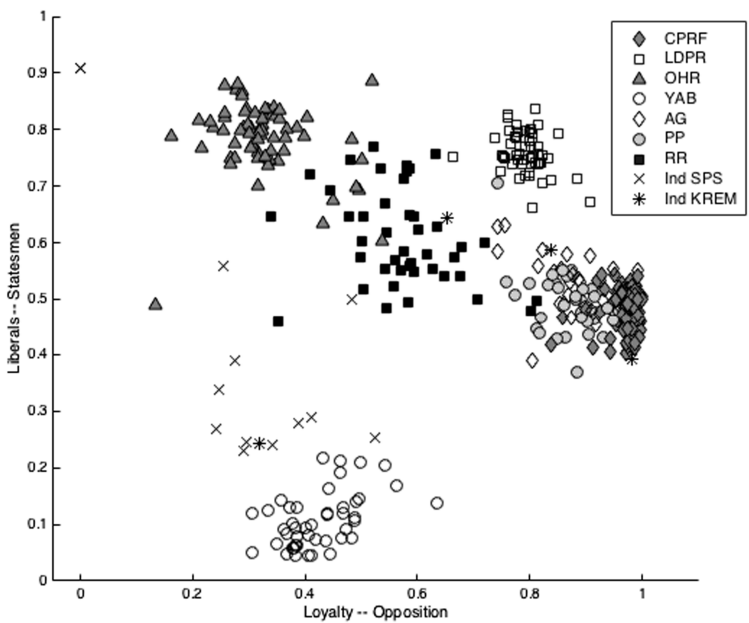

**Fig. 10.** Spatial model of the State Duma in 1996

As Fig. 10–11 demonstrate, the State Duma of the 2<sup>nd</sup> convocation exhibited more pluralism in deputies' positions in 1996 than in 1995. While in 1995 almost all deputy groups were located in the upper right quadrant of the political map, in 1996 loyal to the authorities "Our Home – Russia" and quite moderate "Regions of Russia" and "Yabloko" represented a sort of counterbalance to communists' power in the State Duma. Nevertheless, CPRF and its allies were very numerous which led to the overall opposition spirit towards the President and the Government in the State Duma of the 2<sup>nd</sup> convocation.

In 1996 "Yabloko" had much more liberal (ideology driven) position than in 1995 and was namely the only liberal faction in the State Duma in 1996; thus, variation along Y-coordinate was quite high. Furthermore, it should be noted that in contrast to the State Duma of the 1<sup>st</sup> convocation, the State Duma of the 2<sup>nd</sup> convocation demonstrated better discipline within deputy unions which is reflected in denser structure of clusters on the political map.

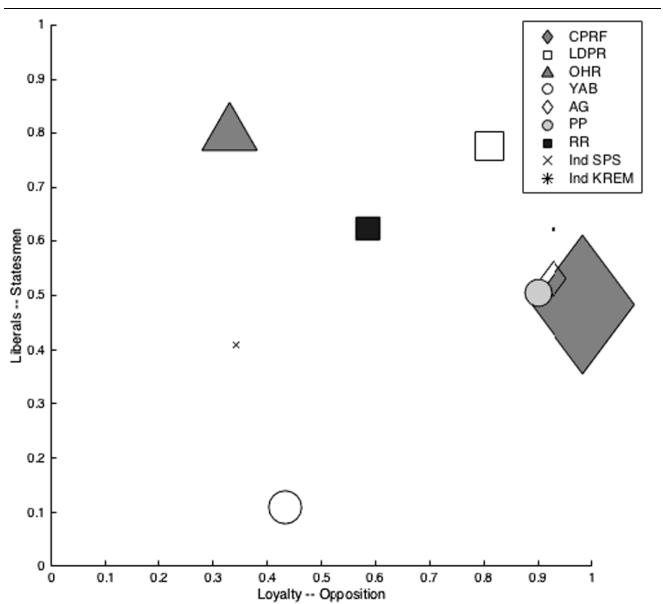

**Fig. 11.** Positions of the deputy unions in the State Duma in 1996

Let us now present the values of the polarization index for the case of the State Duma in 1996

$$Center\ of\ mass = (0.7251, 0.5329);$$
$$P_{Euc} = 0.4154, P_{Man} = 0.3780, P_{Cheb} = 0.5334.$$

The results obtain indicate two main features of the State Duma in 1996. Firstly, it became more polarized (polarization indices increased on average two times when calculated using different metrics) than in 1995. Secondly, on average deputy unions were slightly less opposed to the authorities and significantly more liberal and ideology driven comparing to the case of 1995.[5]

These two features of the State Duma in 1996 can be demonstrated by the following cases. The increased polarization may be supported by an example of the Chairman of the State Duma election. The communist G. Seleznyov was elected to serve as Chairman of the State Duma, notwithstanding "Yabloko" and their allies in the person of independent deputies persistently supported their candidacy, V. Lukin. Overall opposition nature of the Duma is demonstrated by the most notorious case during the 2[nd] convocation, the denunciation of the Belavezha accords in March, 1996. At the same time, the fact that candidacy of V. Chernomyrdin for the post of Prime-Minister was approved by the State Duma supports the results indicating some moderation in deputies' positions and the fact that conflict in the parliament was not too hard.

Let us now present how configuration of deputies changed in 1997 and show what are the consequences of these changes in terms of polarization as well as overall "Loyalty – Opposition" and "Liberals – Statesmen" orientation.

As Fig. 12–13 indicate, the overall configuration of deputies' positions on the political map in 1997 did not change dramatically comparing to the previous year. However, one can observe slight drift to the left of deputy unions "Regions of Russia" and "Yabloko" and Independent deputies which will afterwards form the faction SPS. This can be interpreted as the increased loyalty of this unions towards the executive power. In addition, faction LDPR took more pragmatic, "Statesmen" position in the parliament in 1997.

---

[5] It should be noted that interpretation of the 2[nd] latent factor as "Ideology – Pragmatism" is consistent with "Liberals – Statesmen" interpretation. Liberals tended to be more ideology driven than statesmen, while the latter, in their turn, exhibited more pragmatic willingness to cooperate with executives for the sake of the state interests.

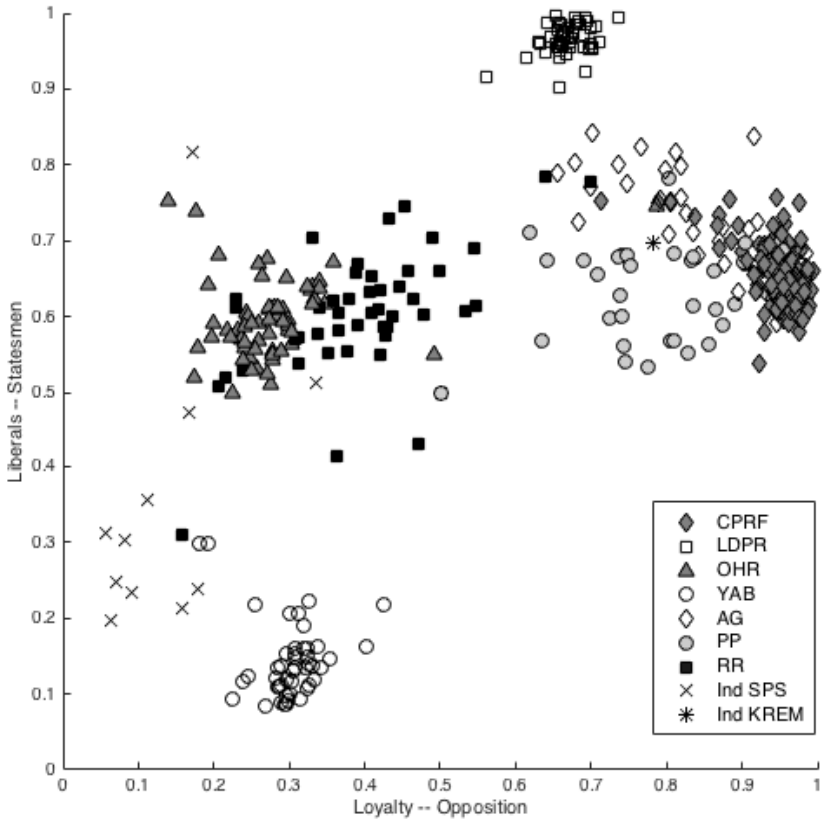

**Fig. 12.** Spatial model of the State Duma in 1997

Several changes in the Government took place in this year. Progovernment candidacies of A. Chubais and B. Nemtsov were appointed for the Deputy Prime Ministers positions. However, several months later the State Duma demanded the President Yeltsin to give A. Chubais the sack.

In general, there was more tension inside the State Duma in 1997 than in previous years.

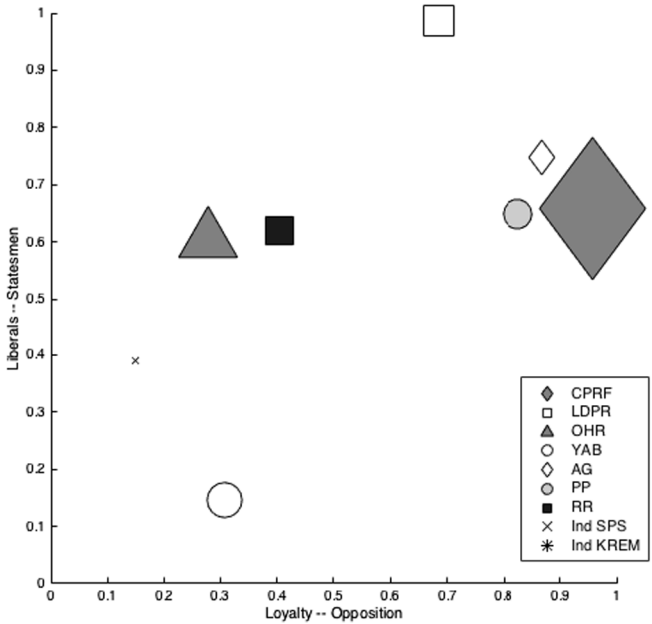

**Fig. 13.** Positions of the deputy unions in the State Duma in 1997

Polarization indices were calculated, and the results obtained are in good agreement with the actual state of affairs.

Polarization index values obtained for the case of 1997:

$Center\ of\ mass = (0.6430, 0.6181);$
$P_{Euc} = 0.4683, P_{Man} = 0.3865, P_{Cheb} = 0.6298.$

In 1998 and 1999 the configuration of the parliament was almost the same. Fig. 14–17 demonstrate specific features of the State Duma in both 1998 and 1999: the deputies' positions highly varied along "Loyalty – Opposition" dimension while remained moderate and almost constant along

"Liberal – Statesmen" dimension with the exception of liberal "Yabloko" and "Statesmen" LDPR.

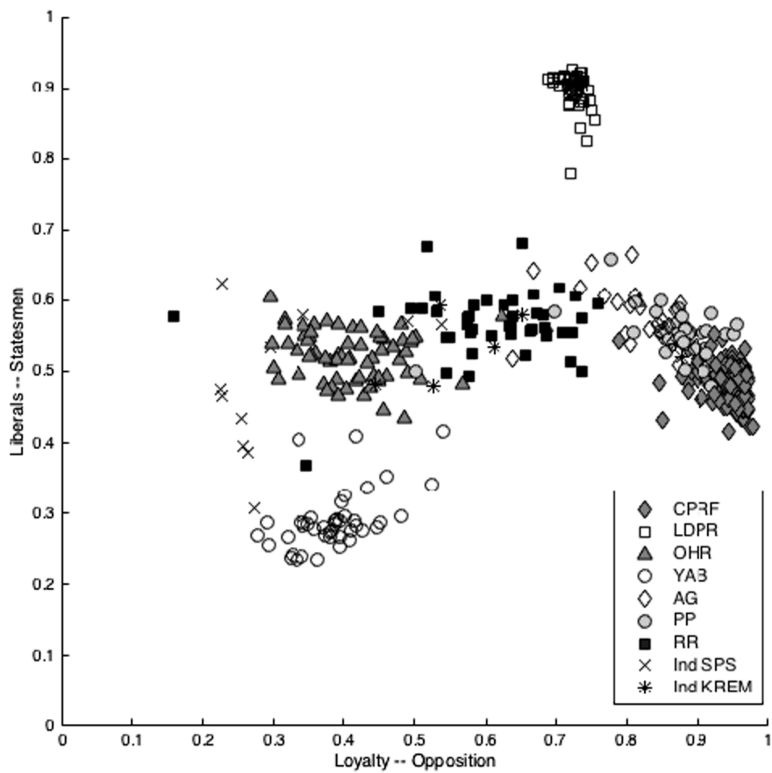

**Fig. 14.** Spatial model of the State Duma in 1998

Polarization index values obtained for the case of 1998:

$Center\ of\ mass = (0.7121, 0.5378);$
$P_{Euc} = 0.3563, P_{Man} = 0.2920, P_{Cheb} = 0.4821.$

Polarization index values obtained for the case of 1999:

$Center\ of\ mass = (0.6922, 0.6160);$
$P_{Euc} = 0.3442, P_{Man} = 0.2913, P_{Cheb} = 0.4602.$

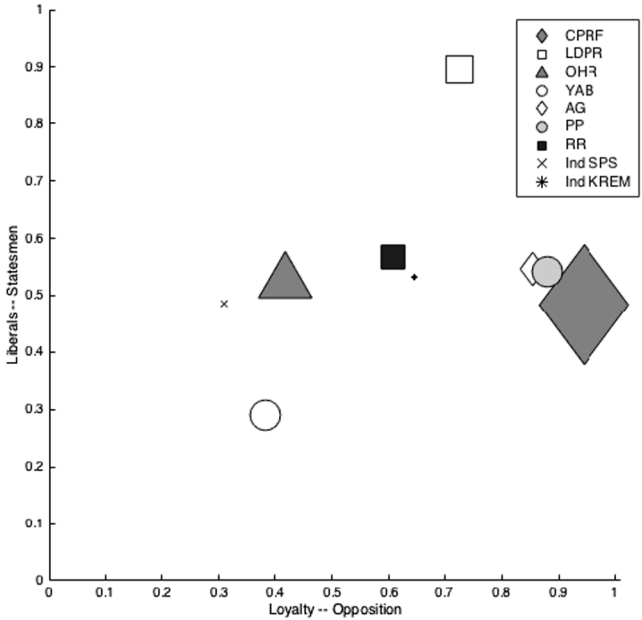

**Fig. 15.** Positions of the deputy unions in the State Duma in 1998

As there was virtually no difference between configurations of deputies' positions in the State Duma in 1998 and 1999, the center of mass which characterize the overall nature of the parliament and the values of polarization index had hardly changed.

On average, the parliament in 1998 and 1999 revealed increased opposition to the current authorities and continued to take moderate position along "Liberals – Statesmen" dimension. At the same time, polarization in the State Duma decreased comparing to the beginning of the convocation.

In 1998 and 1999 the confrontation between the parliament, on the one hand, and the President and the Government, on the other hand, increased. New personnel changes in the Government in 1998 (started with resignation of V. Chernomyrdin) led to the escalation of the conflict, the crisis in the Government and, finally, after the economic crisis in the country, to the resignation of the Government. In addition, the State Duma initiated the im-peachment process in 1998 and only new appointment of the Prime Minister did not allow the parliament to muster a constitutional majority of 300 votes in 1999 on the eve of the final voting on the impeachment issue.

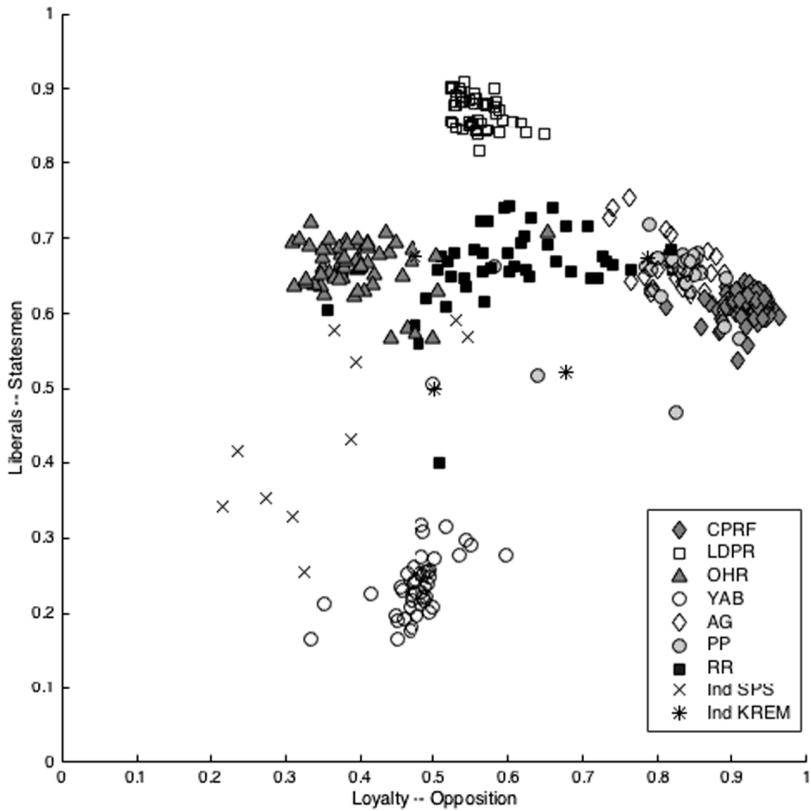

**Fig. 16.** Spatial model of the State Duma in 1999

After the hostiles in Dagestan broke out, the President Yeltsin dismissed the recently formed Government and appointed V. Putin for the post of he Prime Minister simultaneously claiming that Mr. Putin is nominated as his successor.

The decreased values of polarization in the State Duma in 1998 and 1999 is a reflection of increased unity of the deputies in their struggle against executive power (as it was in 1995, at the end of the 1[st] convocation). It is noteworthy, that at the beginning of the convocations the parliament is more polarized which may be treated as a residual effects of tense elections races; however, with the approach of the end of the convocation, the parliament reveals less polarization and increasing opposition towards the President and the Government.

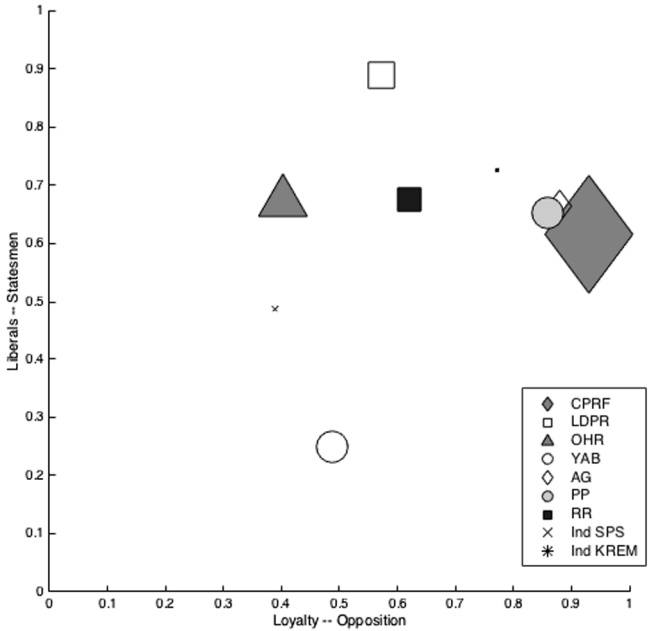

**Fig. 17.** Positions of the deputy unions in the State Duma in 1999

*The State Duma of the 3rd convocation (2000–2003)*

According to the results of the legislative elections held in December, 1999, configuration of forces in the State Duma substantially changed. Recently formed party "Unity" rapidly gained popularity and represented a real counterbalance to the opposition leftists through CPRF in the new convocation.

Deputies in the State Duma of the 3rd convocation formed 9 faction and deputy groups:

- "Agrarian Group" (AG)
- "Unity"

- "Communist Party of the Russian Federation" (CPRF)
- "Liberal Democratic Party of Russia" (LDPR)
- "People's deputy" (MP)
- "Fatherland – all Russia" (FAR)
- "Regions of Russia" (RR)
- "Unity of Right Forces" (SPS)
- "Yabloko" (YAB)

For the State Duma of the 3$^{rd}$ convocation the independent deputies who were not affiliated to any faction or deputy group were considered as the whole entity. The independent deputies whose position was close enough to a particular deputy union were affiliated to this union.

The most numerous factions, opposition CPRF and pro-government "Unity" created two main poles in the newly elected parliament during the first half of its term. However, they both had almost the same moderate positions along "Liberals – Statesmen" dimension. The positions of the next biggest deputy unions, "People's deputy" and "Fatherland – all Russia" were exactly at the center of the unite square during the first half of the 3$^{rd}$ convocation. Fig. 18–21 present configurations of deputies' positions in 2000 and 2001 which were very similar.

As will be demonstrated below, the centers of mass in 2000 and 2001 turned out to be virtually at the same point, and values of polarization index were almost the same which is in accordance with the fact that political maps in 2000 and 2001 look very similar. Let us first present the exact results for both 2000 and 2001 years and then provide a sound interpretation for them.

Polarization index values obtained for the case of 2000

$Center\ of\ mass = (0.4948, 0.5469);$

$P_{Euc} = 0.3722, P_{Man} = 0.3175, P_{Cheb} = 0.4949.$

Polarization index values obtained for the case of 2001

$Center\ of\ mass = (0.5110, 0.5890);$

$P_{Euc} = 0.3857, P_{Man} = 0.3191, P_{Cheb} = 0.5218.$

As the results demonstrate, during the first half of the 3$^{rd}$ convocation, the State Duma was on average much less opposed to the executive power than ever before. At the same time, deputies in general continued to take moderate positions along "Liberals – Statesmen" dimension.

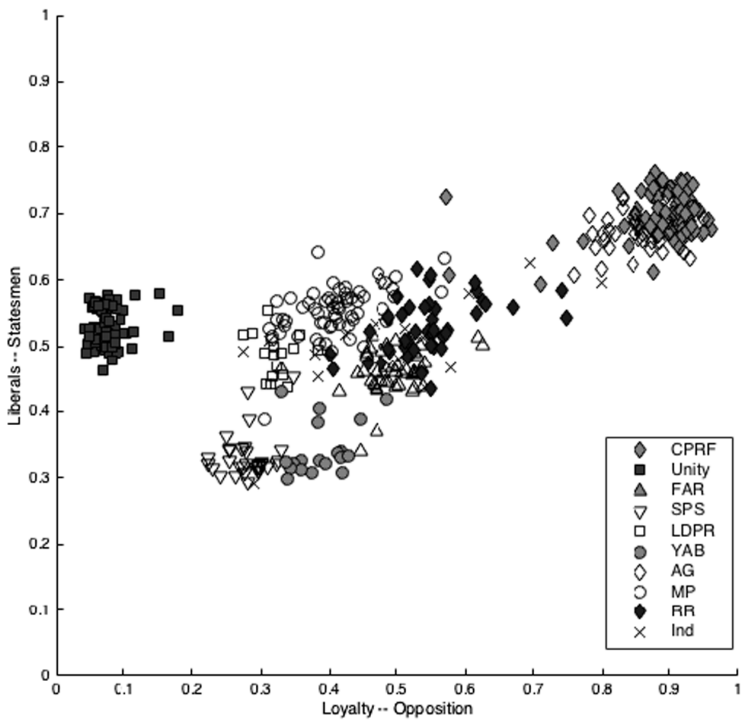

**Fig. 18.** Spatial model of the State Duma in 2000

For the first time the State Duma had really strong counterbalance to the communists; nevertheless, the forces of "Unity" and liberal deputy unions were not enough to form a stable majority loyal to the executive authorities. While during the second half of both the 1st and the 2nd convocation the tension between the legislative and the executive power led to less polarization in the former, with the emergence of strong force loyal to the executives, the confrontation moved inside the parliament. Hence, the parliament on average became less opposed to the executives and more polarized.

The term of the State Duma of the 3rd convocation began with the conflict motivated by portfolio allocation, and during its first half the outcome of voting depended on how flexible the positions of "Unity" and their more moderate colleagues were. For instance, the law on the Federal budget was approved by "Unity", "Yabloko" and SPS without support of the leftists.

At the same time, the constitutional laws on the state symbolics were adopted by "Unity", CPRF, "Fatherland – all Russia" without support of "Yabloko" and SPS which voted against these laws. However, the attempt of "Unity" to compromise with CPRF and their allies did not resolve the conflict and failed to ensure more loyalty from the part of the latter. For example, the law on political parties met heavy resistance from the leftists; moreover, they even initiated the discussion of the question of non-confidence in the Government.

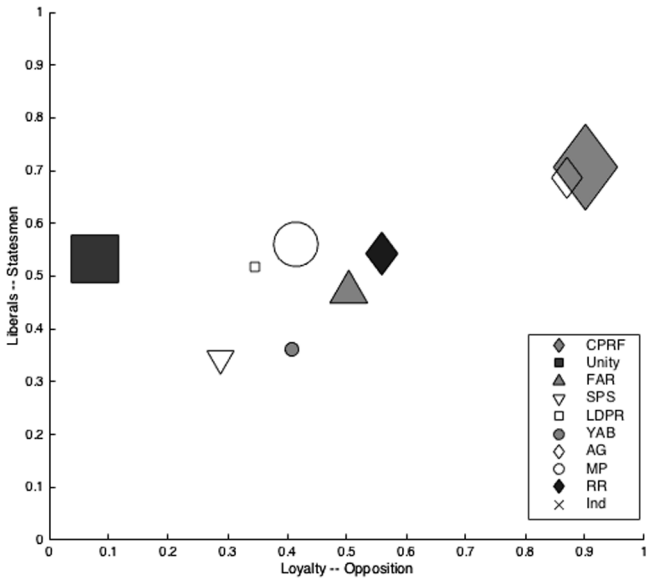

**Fig 19.** Positions of the deputy unions in the State Duma in 2000

The remaining resistance of opposition CPRF and allies notwithstanding, majority loyal to the executives was destined to be formed. In April, 2000 the beginning of merge of "Unity" and "Fatherland – all Russia" and the creation of the unit party on their basis was announced. Moreover, it was also declared that the coalition based on the four centrist deputy unions will be formed. The process aimed at limiting resources of the opposition was launched.

This process and further shift to more loyalty in the parliament was reflected in the configuration of political forces in 2002 and 2003. Fig. 22–25 demonstrate the main changes in political positions of the deputies.

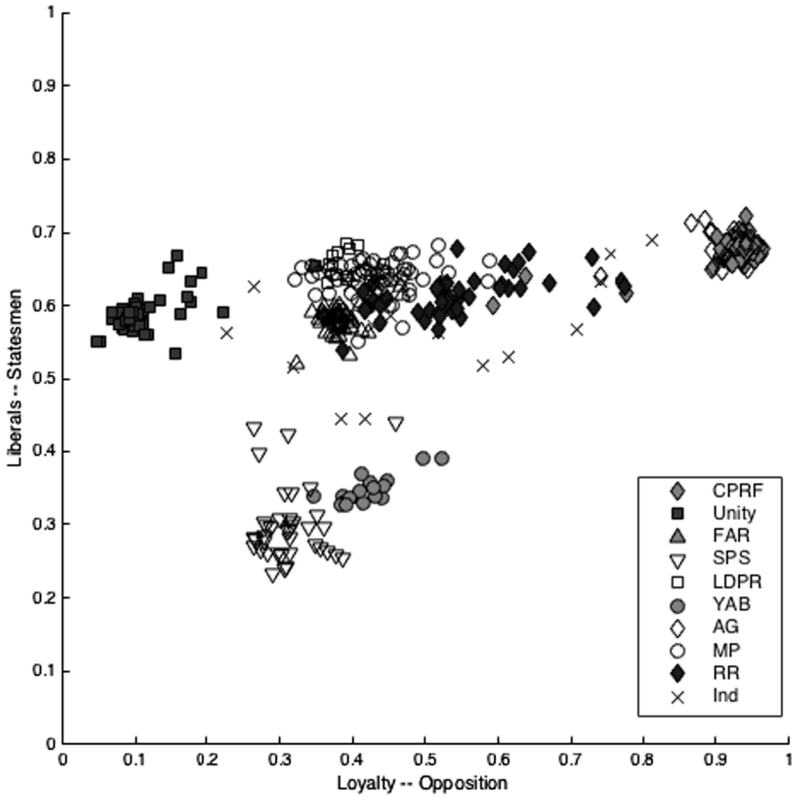

**Fig. 20.** Spatial model of the State Duma in 2001

The consequences of the process of merging of deputy unions "Unity" and "Fatherland – all Russia" reflected in political maps of the State Duma in 2002 and 2003. Positions of deputies affiliated to these unions are virtually the same and extremely loyal to the executive power. Positions of other deputy unions except "Yabloko" shifted to the left revealing more loyalty to the current executives comparing to the first half of the 3[rd] convocation. Thus, two poles represented by "Unity" and allies, on the one hand, and CPRF and allies, on the other hand, became more strongly pronounced. These tendencies reflected in the values of polarization index provided below.

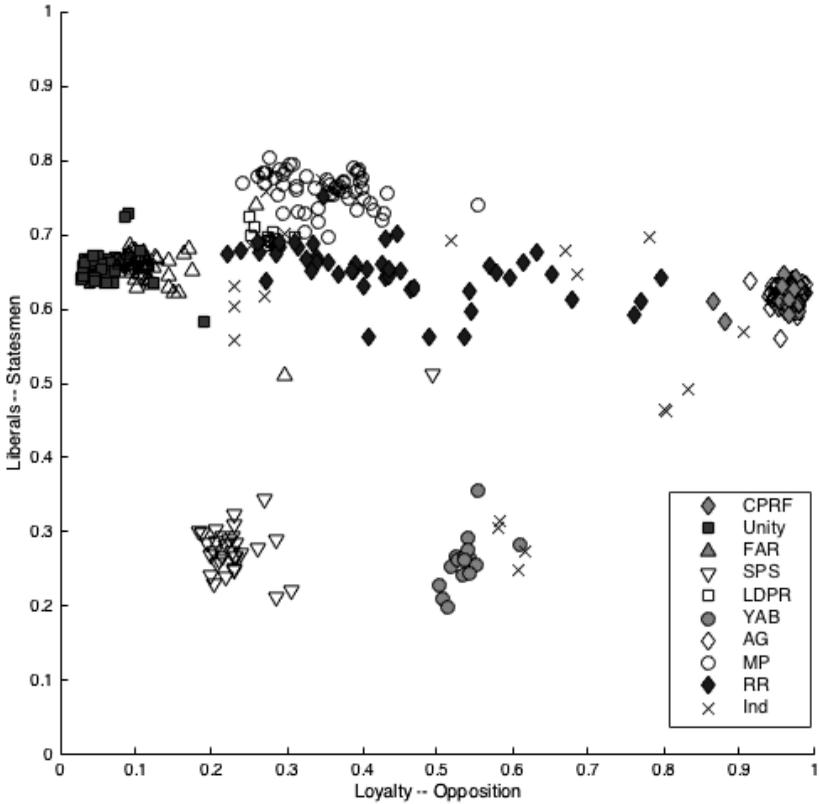

**Fig. 21.** Positions of the deputy unions in the State Duma in 2001

Polarization index values obtained for the case of 2002

$Center\ of\ mass = (0.4564, 0.6124);$
$$P_{Euc} = 0.4805, P_{Man} = 0.3818, P_{Cheb} = 0.6561.$$

Polarization index values obtained for the case of 2003

$Center\ of\ mass = (0.4530, 0.5169);$
$$P_{Euc} = 0.4796, P_{Man} = 0.3797, P_{Cheb} = 0.6654.$$

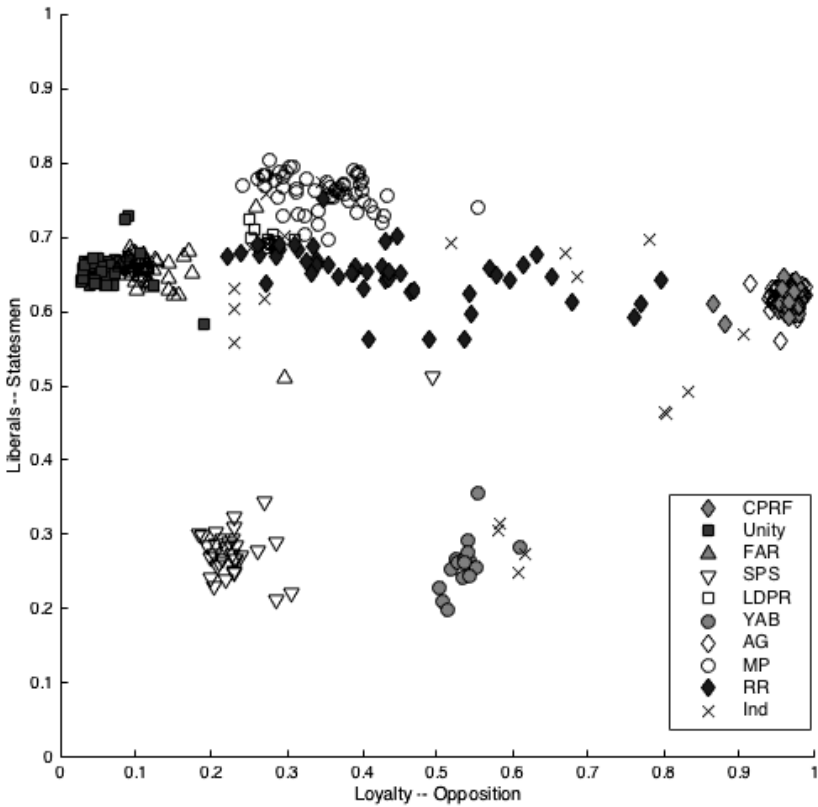

**Fig. 22.** Spatial model of the State Duma in 2002

The coordinates of centers of mass in 2002 and 2003 were very close. They support the fact that during these years the State Duma on the whole exhibited more loyal voting patterns than ever before.

In December, 2002 parliamentary majority was formalized after the inaugural congress of the party "United Russia" formed on the basis of "Unity" and "Fatherland – all Russia". Since then the process of adoption of laws had been accelerated. Resources of the opposition deputy unions became even more limited after the laws on political extremism and holding of a referendum were adopted and some personnel changed in committees were carried out. In 2003 there even was a trend in the parliament according to which only those laws were approved which were supported by recently created loyal majority, whereas position of other factions and deputy groups was ignored. In such a way electricity, housing and communal services, and local government reforms were approved.

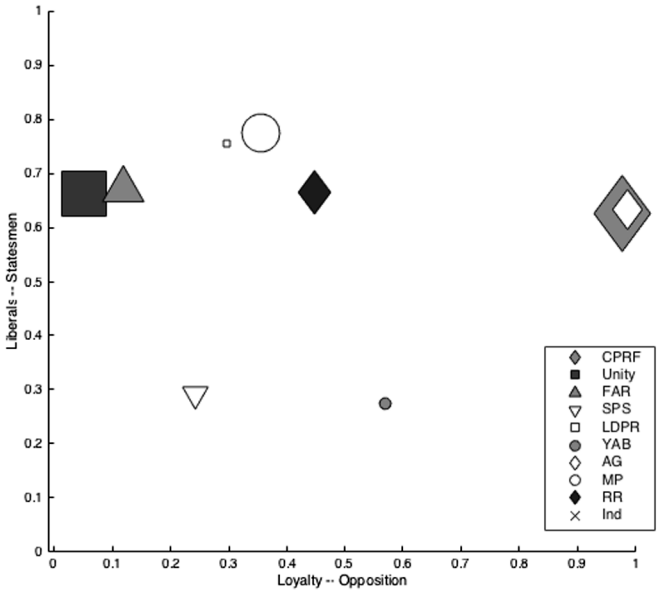

**Fig. 23.** Positions of the deputy unions in the State Duma in 2002

At the same time, during the second half of the term the State Duma of the 3rd convocation was the most polarized Russian parliament throughout the period under consideration. Indeed, the existence of two roughly equally sized poles located at extremely opposed coordinates along "Loyalty – Opposition" dimension which attracted numerous groups with moderate position on the political map consistently led to the significantly high polarization in the State Duma. Opposition factions persistently defended their position and this caused tension inside the parliament.

Below, the summary tables and graphs are provided in order to simplify the comparison.

As Table 3.1 and Fig. 25 show, three versions of the *P* index exhibit the same trend in polarization in the State Duma. The values of *P* index with Chebychev function as a metric are the highest, while values of $P$ index with Manhattan distance are the lowest.

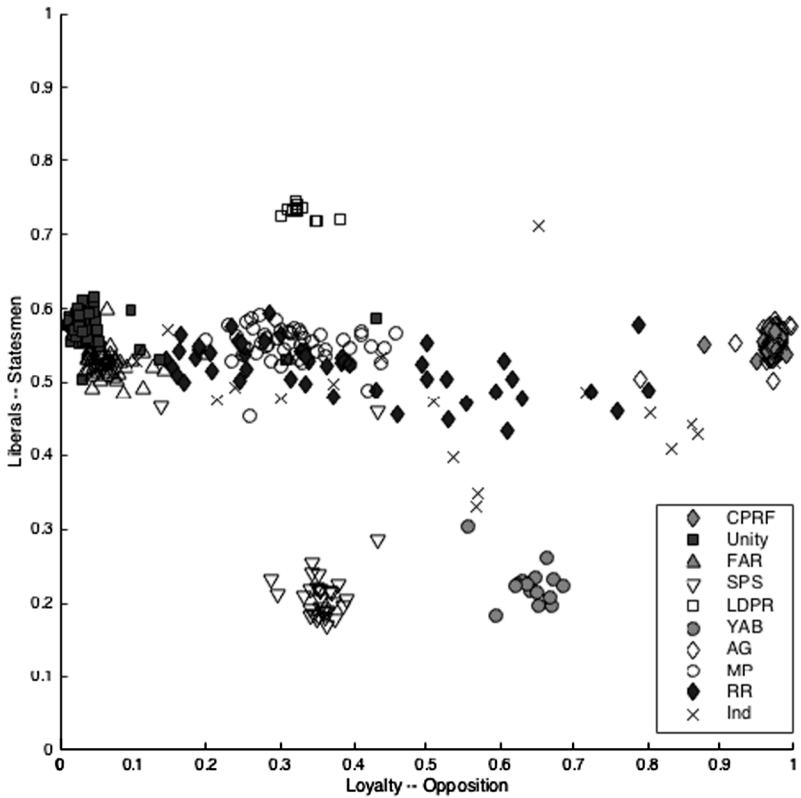

**Fig. 24.** Spatial model of the State Duma in 2003

During the whole period under consideration polarization in the Russian parliament was mostly associated with disagreement of deputies in relation to their loyalty to the current executive authorities. In general, the higher opposition of the whole parliament towards the executives was associated with less polarization, whereas when conflict moved inside the State Duma, and pro-governmental forces were stronger, polarization in the parliament increased.

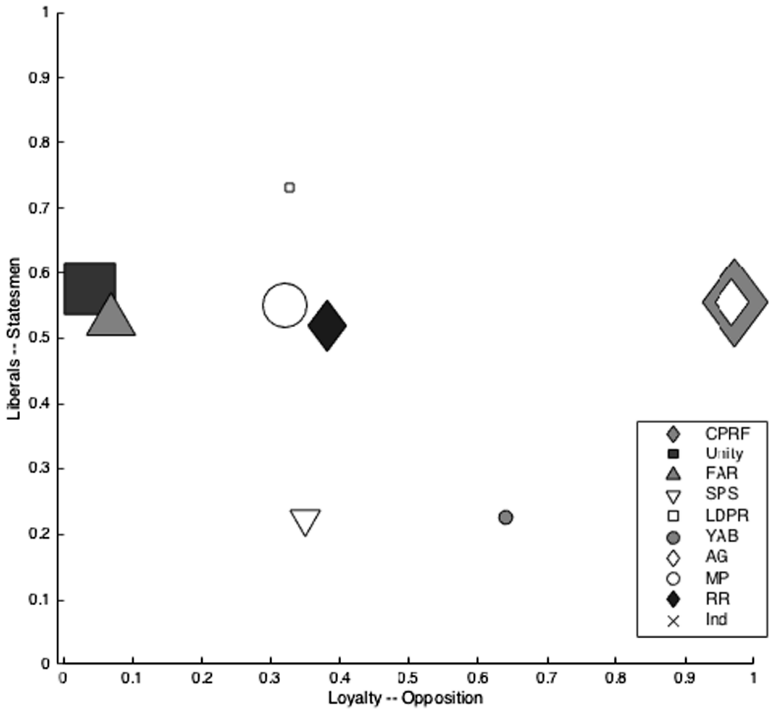

**Fig. 25.** Positions of the deputy unions in the State Duma in 2003

The Russian parliament was the least polarized in 1995 and the most polarized in 2003. Decreased polarization in 1995 matches the onset of the governmental crisis and increased unanimity of deputies in their opposition towards the President and the Government. At the same time, in 1996, when the new convocation of the State Duma was elected, polarization in the parliament increased indicating higher pluralism in opinions and the emergence of progovernmental unions. However, since 1999 polarization in the parliament have been only increasing. During the 3[rd] convocation the forces in the State Duma started to concentrate around two opposed poles represented by progovernmental "Unity" and opposition CPRF. Even though by the end of the term the State Duma of the 3[rd] convocation *de facto* was under control of majority loyal to the executives, the opposition factions were persistent enough in their position which led to relatively high polarization in the parliament.

*Table 1.* Polarization in the State Duma of the 1st – 3rd convocations (groups-points framework)

| Year | Center of mass | $P_{Euc}$ | $P_{Man}$ | $P_{Cheb}$ |
|------|----------------|-----------|-----------|------------|
| 1994 | (0.6746, 0.5523) | 0.3479 | 0.3136 | 0.4487 |
| 1995 | (0.7668, 0.7231) | 0.2190 | 0.1876 | 0.2919 |
| 1996 | (0.7251, 0.5329) | 0.4154 | 0.3780 | 0.5334 |
| 1997 | (0.6430, 0.6181) | 0.4683 | 0.3865 | 0.6298 |
| 1998 | (0.7121, 0.5378) | 0.3563 | 0.2920 | 0.4821 |
| 1999 | (0.6922, 0.6160) | 0.3442 | 0.2913 | 0.4602 |
| 2000 | (0.4948, 0.5469) | 0.3722 | 0.3175 | 0.4949 |
| 2001 | (0.5110, 0.5890) | 0.3857 | 0.3191 | 0.5218 |
| 2002 | (0.4564, 0.6124) | 0.4805 | 0.3818 | 0.6561 |
| 2003 | (0.4530, 0.5169) | 0.4796 | 0.3797 | 0.6654. |

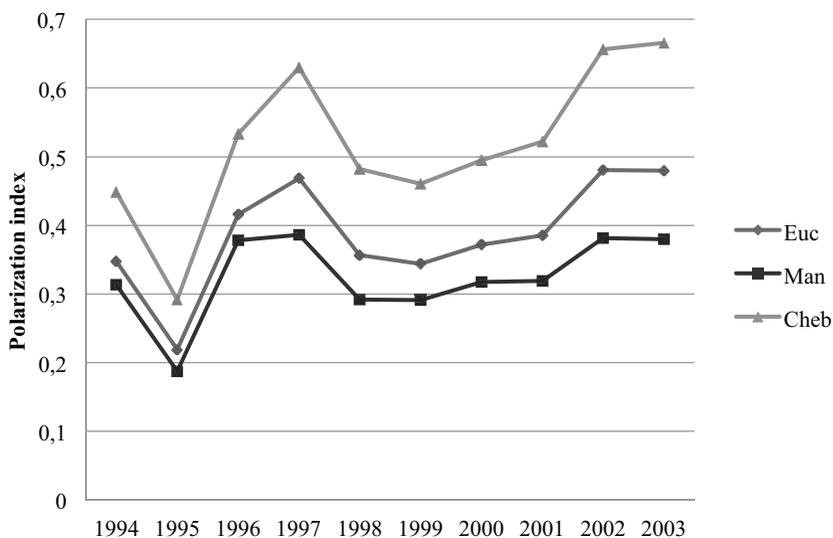

**Fig. 26.** Polarization in the Russian State Duma 1994–2003 (polarization indices estimated within groups-points framework)

# References

*Aleskerov F., Golubenko M.* On the evaluation of a symmetry of political views and polarization of society. Working paper WP7/2003/04. Moscow: State University – Higher School of Economics, 2003. (In Russian.)

*Aleskerov F., Blagoveshchensky N., Satarov G., Sokolova A., Yakuba V.* Power and structural stability in the Russian parliament (in 1905–1917 and 1993–2005). Moscow: Fizmatlit, 2007.

*Atkinson A.* Multidimensional Depriviation: Contrasting Social Welfare and Counting Approaches // Journal of Economic Inequality. 2003. No. 1. P. 51–65.

*Blagoveshchensky N.* Evaluation of the political positions in the State Duma of the 3rd convocation based on the roll-call votes. Moscow: NDEM Foundation, 2004.

*Bossert W., Chakravarty S., D'Ambrosio C.* Multidimensional Poverty and Material Deprivation with Discrete Data // The Review of Income and Wealth. 2013. Vol. 59(1). P. 29–43.

*Chakravarty S., Majumder A.* Inequality, polarization and welfare: Theory and applications // Australian Economic Papers. 2001. No. 40. P. 1–13.

*Chakravarty S., Majumder A., Roy S.* A treatment of absolute indices of polarization // Japanese Economic Review. 2007. No. 58. P. 273–293.

*Chakravarty S., D'Ambrosio C.* Polarization orderings of income distributions // Review of Income and Wealth. 2010. No. 56. P. 47–64.

*Duclos J., Esteban J., Ray D.* Polarization: Concepts, measurement, estimation // Econometrica. 2004. No. 72. P. 1737–1772.

*Esteban J., Ray D.* On the Measurement of Polarization, Boston University / Institute for Economic Development. Working Paper 18, 1991.

*Esteban J., Ray D.* On the measurement of polarization // Econometrica. 1994. Vol. 62(4). P. 819–851.

*Esteban J., Gradin C., Ray D.* An extension of a measure of polarization, with an application to the income distribution of five OECD countries // Journal of Economic Inequality. 2007. No. 5. P. 1–19.

*Esteban J., Ray D.* Polarization, fractionalization and conflict // Journal of Peace Research. 2008. No. 45. P. 163–182.

*Foster J., Wolfson M.* Polarization and the decline of the middle class:

Canada and the U.S. // The Journal of Economic Inequalities. 1994. P. 247–273.

*Gigliarano C., Mosler K.* Constructing indices of multivariate polarization // Journal of Economic Inequality. 2009. No. 7. P. 435–460.

*Gradin C.* Polarizations by subpopulations in Spain, 1973–1991 // Review of Income and Wealth. 2000. No. 48. P. 457–474.

*Horowitz D.* Ethnic groups in conflict. Berkeley: University of California Press, 1985.

*Lipacheva A.* Comparison of Polarization and Bi-Polarization Indices in Some Special Cases. Working paper WP7/2015/06. Moscow: State University – Higher School of Economics, 2015.

*Montalvo J, Reynal-Querol M.* Ethnic polarization, potential conflict and civil wars // American Economic Review. 2005. No. 95. P. 796–816.

*Nolan B., Whelan V.* On the Multidimensionality of Poverty and Social Exclusion. Oxford: Oxford University Press, 2007. P. 146–165.

*Poole K., Rosenthal H.* A Spatial Model for Legislative Roll Call Analysis. GSIA Working Paper 5, 1983. P. 83–84.

*Poole K., Rosenthal H.* The Polarization of American Politics // Journal of Politics. 1984. Vol. 46(4). P. 1061–79.

*Poole K., Rosenthal H.* Congress: A Political-Economic History of Roll Call Voting. Oxford: Oxford University Press, 1997.

*Rodríguez J., Salas R.* Extended bi-polarization and inequality measures // Research on Economic Inequality. 2003. No. 9. P. 69–83.

*Satarov G.* Analysis of political structure of legislative bodies based on roll-call votes // Russian Monitor: Archive of modern policy. 1993. No. 1.

*Scheicher C.* Measuring polarization via poverty and affluence // Discussion Papers in Statistics and Econometrics / Seminar of Economic and Social Statistics. University of Cologne, 2010. No. 3.

*Wang Y., Tsui K.* Polarization orderings and new classes of polarization indices // Journal of Public Economic Theory. 2000. P. 349–363.

*Wolfson M.* When inequalities diverge // The American Economic Review. 1994. No. 48. P. 353–358.

*Wolfson M.* Divergent inequalities: Theory and empirical results // Review of Income and Wealth. 1997. No. 43. P. 401–421.

*Zhang X., Kanbur R.* What difference do polarization measures make? An application to China // The Journal of Development Studies. 2001. No. 37. P. 85–98.

INDEM-Statistic, information-analytical project of INDEM Foundation <http://www.indem.ru/indemstat/index.htm>.

VoteView Website by Poole and Rosenthal <http://voteview.com>.

Разработан многомерный индекс поляризованности на основе индекса Алескерова – Голубенко. Представлено несколько версий индекса поляризованности в зависимости от разных функций расстояния. Рассмотрены основные свойства индекса. Численно изучено поведение индекса поляризованности в случае «равномерного» распределения групп в единичном квадрате и единичном трехмерном кубе.

Исследована поляризованность в Государственной думе Российской Федерации (1994–2003 гг.) с помощью построенного индекса. При анализе использована ранее разработанная на основании поименных голосований депутатов двумерная модель Российской Государственной думы. Результаты применения многомерного индекса поляризованности согласуются с соответствующими политическими событиями. Показано, что поляризованность в Государственной думе была сопряжена главным образом со степенью напряженности в ее взаимоотношениях с исполнительной властью. В частности, чем более выраженной была конфронтация между законодательной и исполнительной ветвями власти, тем менее поляризованной была Государственная Дума, и наоборот.

Ключевые слова: поляризованность, индекс поляризованности, многомерный индекс поляризованности, индекс Алескерова – Голубенко, поляризованность в Государственной думе Российской Федерации

Алескеров Фуад Тагиевич, Олейник Виктория Валерьевна

**Многомерный индекс поляризованности
и его применение к анализу Государственной думы
Российской Федерации (1994–2003 гг.)**

(*на английском языке*)

Зав. редакцией оперативного выпуска *А.В. Заиченко*
Технический редактор *Ю.Н. Петрина*